\documentclass[12pt,ws-ijmpa]{article}

\usepackage{latexsym,amsmath,amssymb}

\newcommand{\be}{\begin{equation}}
\newcommand{\ee}{\end{equation}}
\newcommand{\beu}{\begin{equation*}}
\newcommand{\eeu}{\end{equation*}}
\newcommand{\bea}{\begin{eqnarray}}
\newcommand{\eea}{\end{eqnarray}}
\newcommand{\beaa}{\begin{eqnarray*}}
\newcommand{\eeaa}{\end{eqnarray*}}
\newcommand{\cR}{\check{R}}

\newcommand{\g}{{\mathfrak g}}
\newcommand{\h}{{\mathfrak h}}
\newcommand{\m}{{\mathfrak m}}

\advance\textheight by \topskip
\newdimen\tableauside\tableauside=1.0ex
\newdimen\tableaurule\tableaurule=0.4pt
\newdimen\tableaustep
\def\phantomhrule#1{\hbox{\vbox to0pt{\hrule height\tableaurule width#1\vss}}}
\def\phantomvrule#1{\vbox{\hbox to0pt{\vrule width\tableaurule height#1\hss}}}
\def\sqr{\vbox{%
  \phantomhrule\tableaustep
  \hbox{\phantomvrule\tableaustep\kern\tableaustep\phantomvrule\tableaustep}%
  \hbox{\vbox{\phantomhrule\tableauside}\kern-\tableaurule}}}
\def\squares#1{\hbox{\count0=#1\noindent\loop\sqr
  \advance\count0 by-1 \ifnum\count0>0\repeat}}
\def\tableau#1{\vcenter{\offinterlineskip
  \tableaustep=\tableauside\advance\tableaustep by-\tableaurule
  \kern\normallineskip\hbox
    {\kern\normallineskip\vbox
      {\gettableau#1 0 }%
     \kern\normallineskip\kern\tableaurule}%
  \kern\normallineskip\kern\tableaurule}}
\def\gettableau#1 {\ifnum#1=0\let\next=\null\else
  \squares{#1}\let\next=\gettableau\fi\next}

\begin{document}
\baselineskip 17pt
\parindent 8pt
\parskip 12pt

%\begin{flushright}
%\end{flushright}
\hfill ESI 1514

\begin{center}
{\Large {\bf  Introduction to Yangian symmetry\\[0.05in]
 in integrable field theory}}\\
\vspace{0.5cm} {\large N. J. MacKay\footnote{\tt
nm15@york.ac.uk}}\\ {\em Department of Mathematics, University of
York,\\
Heslington Lane, York YO10 5DD, UK}
\end{center}

\vskip 0.2in
 \centerline{\small\bf ABSTRACT}
\centerline{
\parbox[t]{5in}{\small An introduction to Yangians and their
representations, to Yangian symmetry in 1+1D integrable (bulk)
field theory, and to the effect of a boundary on this symmetry.
\noindent }}

\vskip 0.4in
\section{Introduction}

The Yangian $Y(\g)$ of a simple Lie algebra $\g$ was introduced by
Drinfeld in 1985-6 \cite{drinf1,drinf2}, emerging naturally from
the combination of $\g$-symmetry with integrability in 1+1D
models. Yangians tended initially to be overshadowed by
$q$-deformed algebras, in which a major industry developed.
However, in 1990 Cherednik presciently wrote `I think that
[Yangians] should be more important for mathematics and physics
than the $q$-analogues of universal ebveloping algebras now in
common use' \cite{cher}, and perhaps that is now becoming more
widely believed. Certainly Yangians are making an appearance on
both sides of the gauge/string correspondence \cite{DNW,string}.

The intention of these notes is to give a pedagogical introduction
for physicists. They complement the article by Bernard
\cite{bern92} and the reviews for mathematicians by Molev
\cite{molev} and in the book by Chari \& Pressley \cite{CPbook}.
Section two is mathematical: its aim is to introduce some of the
nice properties of $Y(\g)$, including, in the last subsection, one
salient point in which it behaves rather more nicely than $\g$.
Section three explains how $Y(\g)$ emerges in 1+1D field theory,
both classical and quantum (we shall not be discussing lattice
models or spin chains here). In the final section we explain what
happens to $Y(\g)$ in the presence of a boundary. \vfill\pagebreak

\section{Yangians and their representations}

\subsection{Yangians}

\subsubsection{Definition of $Y(\g)$: algebra and co-algebra}

Let the simple Lie algebra $\g$  be generated\footnote{To avoid
becoming mired in detail, we have used anti-Hermitian generators
(hence no `$i$'), and, more importantly, compactness (hence an
inner product proportional to $\delta_{ab}$, so that we won't have
to distinguish `up' from `down' indices).} by $\{I_a\}$,
$a=1,\ldots,$dim$\,\g$, with structure constants $f_{abc}$,
\be\label{Y1}\left[ I_a , I_b \right] = f_{abc} I_c \,,\ee
 and (trivial) coproduct\footnote{The enveloping algebra $U\g$ of $\g$
consists of (powers, polynomials and) series in the $I_a$ subject
to the Lie bracket, regarded as a commutator.} $\Delta:
U\g\rightarrow U\g\otimes U\g$, \be\label{Y1del}
 \Delta(I_a) = I_a \otimes 1
+ 1 \otimes I_a \,. \ee
 For those new to it, the
coproduct is a generalization of the usual rule for addition of
spin. As such, the principal constraints on $\Delta$ are that it
be coassociative, \be\label{coassoc} (\Delta\otimes
1)\Delta(x)=(1\otimes\Delta)\Delta(x)\ee for all $x\in\g$ (so that
the action of $x$ on a 3-particle state is unique), and that it
 be a homomorphism, \be\label{hom} \Delta([x,y])=[\Delta(x),\Delta(y)]
  \ee for all $x,y\in\g$ (so that multiparticle states carry
 representations of the symmetry algebra). As we shall see below,
 there are non-trivial ways in which this can be achieved.

The Yangian \cite{drinf1,drinf2} $Y(\g)$ is the enveloping algebra
generated by these and a second set of generators $\{J_a\}$, in
the adjoint representation of $\g$ so that \be\label{Y2} \left[
I_a , J_b \right] =   f_{abc} J_c \,, \ee but with a non-trivial
coproduct \be\label{Y2del} \Delta:Y(\g)\rightarrow Y(\g)\otimes
Y(\g)\,,\qquad \Delta(J_a) = J_a \otimes 1 + 1 \otimes J_a +
{\alpha\over 2}f_{abc} I_c \otimes I_b\, \ee for a parameter
$\alpha\in{\mathbb C}$. Note that (\ref{coassoc},\,\ref{hom}) hold
for all the $I_a,J_a$.

The commutator $[J_{a},J_{b}]$ is not fully specified, but is
constrained by the requirement that $\Delta$ be a homomorphism (as
explained in the next subsection): \be\label{Y3}
[J_a,[J_b,I_c]]-[I_a,[J_b,J_c]] = \alpha^2
a_{abcdeg}\{I_d,I_e,I_g\} \,, \ee where \be
 a_{abcdeg}={1\over{24}}
 f_{adi} f_{bej}f_{cgk}f_{ijk}
 \hspace{0.1in},\hspace{0.15in}\{x_1,x_2,x_3\}=
 \sum_{i\neq j\neq k}x_{i}x_{j}x_{k} \,,
\ee and \be\label{Y4} [[J_a,J_b],[I_l,J_m]] +
[[J_l,J_m],[I_a,J_b]] = \alpha^2 \left( a_{abcdeg}f_{lmc} +
a_{lmcdeg}f_{abc} \right) \left\{ I_d,I_e,J_g \right\} \,. \ee For
$\g=a_1$,  (\ref{Y3}) is trivial, while for $\g\neq a_1$,
(\ref{Y3}) implies (\ref{Y4}), which is thus
redundant\footnote{Throughout these lectures, we write specific
$\g$ as $a_n={\mathfrak{su}}_{n+1}$, $b_n={\mathfrak{so}}_{2n+1}$,
$c_n={\mathfrak{sp}}_n$, $d_n={\mathfrak{so}}_{2n}$, along with
$e_6,e_7,e_8,f_4,g_2$.}.

In the original sense of the word `quantum' in `quantum group',
the parameter $\alpha$ is proportional to $\hbar$: it measures the
deformation of the `auxiliary' Lie algebra required to make the
quantum inverse scattering method work. In the next lecture, by
contrast, we shall see $Y(\g)$ appearing explicitly as a charge
algebra, with $\alpha=1$ and $\hbar$ making a conventional
appearance on the right-hand side of each commutator.

Finally, there are also other structures on which we place less
emphasis but which make $Y(\g)$ a Hopf algebra and which we give
for completeness: a co-unit \be \epsilon: Y(\g)\rightarrow
{\mathbb C}\,,\qquad \epsilon (I_a)=0,\quad\epsilon(J_a)=0\ee
(physically a one-dimensional vacuum representation, trivial for
$Y(\g)$), and an antipode \be \label{antipode}s:Y(\g)\rightarrow
Y(\g)\,,\qquad s(I_a)=-I_a,\quad s(J_a)=-J_a+{1\over 2}f_{abc}I_c
I_b\,,\ee an anti-automorphism (and physically a
$PT$-transformation).

\subsubsection{Drinfeld's `terrific' relation}

 Drinfeld
called the relations (\ref{Y3}) and (\ref{Y4}) `terrific'
\cite{drinf2}, and it is worth explaining their origin and
significance further (see also \cite{gron}). First, the left-hand
side of (\ref{Y3}) is a little more intuitive if we instead write
it as \be\label{Y3'} [J_a,[J_b,I_c]]-[I_a,[J_b,J_c]] =
f_{d(ab}[J_{c)},J_d], \ee where $(abc)$ means `the sum of $abc$
and cycles thereof'. One way of viewing $Y(\g)$ is as a
deformation of the polynomial algebra $\g[z]$: if $\alpha=0$, then
the algebra reduces to that of $I_a$ and $J_a\equiv z I_a$,
whereupon (\ref{Y3}) is just the Jacobi identity
$f_{d(ab}f_{c)de}=0$. So a natural way to think about $Y(\g)$ is
as a graded algebra, in which $I_a$ has grade zero and $J_a$ and
$\alpha$ each grade one, and (\ref{Y3}) and (\ref{Y4}) are viewed
as constraints on the construction of higher-grade elements. For
example, suppose we define a grade-two element \be K_a \equiv
{1\over c_A} f_{abc}[J_c,J_b] \qquad{\rm (where\;\;}
f_{abc}f_{dcb}=c_A \delta_{ad}\;). \ee Then if we write\be
[J_b,J_c]=f_{bcd}K_d + X_{bc}\,, \ee (\ref{Y3}) fixes $X_{bc}$:
for suppose not, that there exists another possible $X'_{bc}$.
Then, setting $Y_{bc}=X'_{bc}-X_{bc}$, (\ref{Y3}) implies that
$f_{d(ab}Y_{c)d}=0$, and thus (equivalent to the statement
 that the second cohomology $H^2(\g)=0$) that $Y_{cd}=f_{cde}Z_e$
 for some $Z_e$. But then
$c_A Z_a = f_{abc} f_{cbd}Z_d = f_{abc}Y_{cb}=c_A(K_a-K_a)=0$, so
$Z_a=0$ and so $X'_{bc}=X_{bc}$.

The origin of (\ref{Y3}) lies in first postulating $\Delta(J_a)$
in (\ref{Y2del}) and then requiring that this be a homomorphism.
To see this, first let $u_{ab}$ be such that $u_{ab}=-u_{ba}$ and
\be\label{u1} u_{ab} [ I_a,I_b ] = 0 \,. \ee Now compute \be
u_{ab}\left(\Delta \left( [J_{a},J_{b}] \right) - 1\otimes
[J_a,J_b] - [J_a,J_b] \otimes 1 \right) \,. \ee The parts of this
expression involving $J$ disappear because of (\ref{u1}), whilst
the remainder is
\begin{equation}\label{u2}
{1\over 2}u_{ab}f_{ade}f_{bgh}f_{dgk} \left( I_k \otimes I_e I_h +
I_e I_h\otimes I_k \right) \,.
\end{equation}
Because of (\ref{u1}), or $f_{abc}u_{ab}=0$, we may write \be
u_{ab}= v_{dea} f_{deb} - v_{deb} f_{dea}   \ee (equivalent to the
second homology $H_2(\g)=0$). Requiring $\Delta$ to be a
homomorphism for all $v$, and using the Jacobi identity twice, we
obtain (\ref{Y3}).

\subsubsection{The $R$-matrix}

$Y(\g)$ is closely related to the Yang-Baxter equation, which has
a rich literature in its own right (see \cite{jimbo}). A nice way
to see this, of which we give a sketch here, is to define a new
object, the monodromy matrix, \be T(\lambda) \equiv \exp\left(
-{1\over \lambda}t^aI_a + {1\over \lambda^2}t^aJ_a-{1\over
\lambda^3}t^aK_a+\ldots\right), \ee where $\lambda\in{\mathbb C}$
is a new, `spectral' parameter. The $\ldots$ denote higher terms,
of an appropriate-grade element constructed by repeated
commutation of $J$s, and the $t^a$ are a second set of generators
of $\g$ (commuting with $Y(\g)$), to be thought of as matrices
(perhaps in the defining representation of $\g$, where this
exists) with elements $t^a_{ij}$. Thus, overall, $T$ is a matrix,
with entries which are elements of $Y(\g)$.

The significance of $T$ lies in the fact that \be\label{Tdel}
\Delta(T_{ij}(\lambda))=T_{ik}(\lambda) \otimes
T_{kj}(\lambda)\,.\ee The first few terms are easily checked using
(\ref{Y1del},\,\ref{Y2del}): at order $\lambda^{-2}$, for example,
the non-trivial terms in the $\Delta(J_a)$ on the left are matched
on the right not only with the order-$\lambda^{-2}$ terms in each
exponential but also with the cross-terms from multiplying the
order-$\lambda^{-1}$ term in each $T$.

Now, $Y(\g)$ has an (outer) automorphism \be\label{auto}
L_\mu\,:\quad I_a\mapsto I_a\,,\qquad J_a\mapsto J_a+\mu I_a
\qquad(\mu\in{\mathbb C})\ee (equivalent to $z\mapsto z+\mu$ in
the polynomial algebra if $\alpha=0$), whose action on $T$ is \be
T(\lambda)\mapsto T(\lambda+\mu)\,.\ee Let us consider the
intertwiners $\cR$, which are required to satisfy \be\label{R}
\cR(\nu-\mu)\,. \;L_\mu\!\times\! L_\nu\left(\Delta(x)\right)
=L_\nu\!\times\! L_\mu\left(\Delta(x)\right) .\,\cR(\nu-\mu)\ee
for any $x\in Y(\g)$. (Strictly, we should only take
representations of this, and our intertwiner, conventionally
written $\cR$, and then an `$R$-matrix', is often written
$\cR={\bf P}R$ where ${\bf P}$ permutes the two module elements in
the tensor product---but we wish to defer all discussion of
representations to the next section.)

Thus \be \label{RTT}\cR(\nu-\mu)\, T_{ik}(\mu)\otimes T_{kj}(\nu)
= T_{ik}(\nu)\otimes T_{kj}(\mu) \,\cR(\nu-\mu)\,.\ee  There are
then two maps \be T_{ik}(\mu)\otimes T_{kl}(\nu)\otimes
T_{lj}(\lambda) \mapsto T_{ik}(\lambda)\otimes T_{kl}(\nu)\otimes
T_{lj}(\mu)\,,\ee and their equivalence \be\label{YBE}
\cR(\lambda-\nu)\otimes 1\,.\, 1\otimes \cR(\lambda-\mu)\,.\,
\cR(\nu-\mu)\otimes 1 = 1\otimes\cR(\nu-\mu)\,.\,
\cR(\lambda-\mu)\otimes 1\,.\,1\otimes \cR(\lambda-\nu)\ee is the
Yang-Baxter equation (YBE), illustrated schematically in fig.1.

This equation is familiar from 1+1D $S$-matrix theory, where it is
the condition for consistent factorization of the multiparticle
$S$-matrix into two-particle factors.
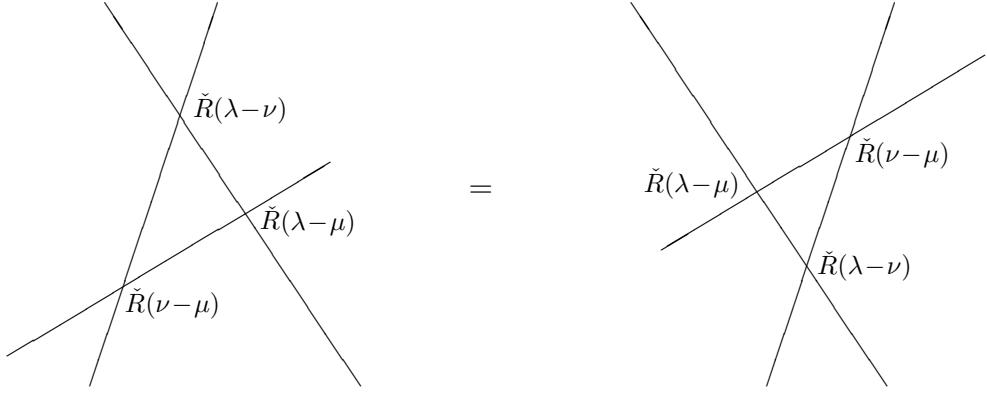
\begin{figure}[htb]
\hspace{32mm} \setlength{\unitlength}{1mm}
\begin{picture}(120,30)(15,15)
\put(8,-6){\line(1,3){17}} \put(98,-6){\line(1,3){17}}
\put(44,-6){\line(-2,3){34}} \put(114,-6){\line(-2,3){34}}
\put(-3,-2){\line(5,3){43}} \put(127,38){\line(-5,-3){43}}
\put(60,20){\makebox(0,0){=}}
\put(19,5){\makebox(0,0){\footnotesize $\cR(\nu\!-\!\mu)$}}
\put(37,16){\makebox(0,0){\footnotesize$\cR(\lambda\!-\!\mu)$}}
\put(28,31){\makebox(0,0){\footnotesize$\cR(\lambda\!-\!\nu)$}}
\put(111,10){\makebox(0,0){\footnotesize$\cR(\lambda\!-\!\nu)$}}
\put(88,21){\makebox(0,0){\footnotesize$\cR(\lambda\!-\!\mu)$}}
\put(116,25){\makebox(0,0){\footnotesize$\cR(\nu\!-\!\mu)$}}
\end{picture}\vspace{20mm}
\caption{The Yang-Baxter equation}
\end{figure}
Each line in the figure will carry a representation of $Y(\g)$.
The simplest case, in which this is ${\mathbb C}^2$, yielded the
first solution of this equation, due to Yang \cite{Yang},
\be\label{firstR}
\cR(\mu) = \left(\begin{array}{cccc} 1+\mu &  0 & 0 & 0\\
                      0 & 1 & \mu & 0\\
                       0& \mu & 1 & 0\\
                  0  &0 & 0& 1+\mu
\end{array}\right) \,.
\ee $Y({\frak sl}_2)$ can be built from it, and it was in honour
of this that Drinfeld named the Yangian.

 A theorem of Belavin and Drinfeld \cite{beldr} is that
(subject to certain technical conditions) as
$\mu\rightarrow\infty$ all YBE solutions which are rational
functions of $\mu$ (and we shall see that this is so for the
Yangian $\cR(\mu)$ in the next section) are of the asymptotic form
\be \cR(\mu) = {\bf P}\left( 1\otimes 1 + {1\over \mu} I_a\otimes
I_a + {\cal O}(\mu^{-2})\right),\ee This leads not only to the
uniqueness of $Y(\g)$, but also to the possibility of
rediscovering much about Lie algebras and their representations
purely by studying YBE solutions \cite{cher2}.

\subsection{Representations of Yangians}

Since $Y(\g)\supset \g$, and representations (`reps') of $Y(\g)$
will also be reps of $\g$, the representation theory of $\g$ is a
good starting point. Recall that, for a Lie algebra $\g$ of rank
$r$, there are $r$ distinguished, `fundamental' irreducible
representations (`irreps'). The story is similar for $Y(\g)$,
which also has $r$ fundamental (and finite-dimensional) irreps
\cite{gron,drinf88}.

However, a rep which is  $Y(\g)$-irreducible may be
$\g$-reducible, and this is typically the case for the fundamental
irreps of $Y(\g)$, whose $\g$-components are the corresponding
fundamental irrep of $\g$ and (generally) some others. These
decompositions appeared incrementally in the literature
\cite{ogiev86,chari91,KR}; for a full enumeration for simply-laced
$\g$ see \cite{kleber96}.

\subsubsection{$Y(\g)$-reps which are $\g$-irreps}

The simplest situation is clearly when a $\g$-irrep is extensible
to a $Y(\g)$-irrep, and there are no other components. Drinfeld
enumerated the cases for which this occurs \cite{drinf1}. Starting
from an irrep $\rho$ of $\g$, he constructed a rep $\tilde{\rho}$
of $Y(\g)$ by setting
\begin{equation}\label{rep}
\tilde\rho(I_a) = \rho(I_a) \,, \qquad \tilde\rho(J_a) = 0 \,.
\end{equation}
Now, although $\tilde\rho$ is clearly consistent with
(\ref{Y1},\,\ref{Y2}), it is not, in general, consistent with
(\ref{Y3}) (we specialize here to $\g\neq a_1$, and so do not
consider (\ref{Y4}) separately). Consistency is only possible for
irreps in which the right hand side of (\ref{Y3}) vanishes. This
is the case for the following irreps.

 Let $n_i$ be the coefficient of the simple
root ${\alpha}_i\in{\mathbb R}^r$ ($i=1,\ldots,r$) in the
expansion of the highest root ${\alpha}_{\rm max}$ of $\g$, and
let $k_i = ({\alpha}_{\rm max},{\alpha}_{\rm
max})/({\alpha}_i,{\alpha}_i)$. Let the corresponding fundamental
weight be ${\omega}_i$. The irrep of $\g$ with highest weight
${\Omega}$ may then be extended to an irrep of $Y(\g)$ using
(\ref{rep}) for \be
\begin{array}{rcl}
 & (i) & {\Omega}={\omega}_i \quad {\rm when } \quad
 n_i=k_i\\[0.2in]

 {\rm and} & (ii) &
  {\Omega}=t{\omega}_i\quad {\rm when }
 \quad n_i=1
 \qquad (t\in{\mathbb Z}).\end{array}\label{irreps} \ee
These include all the fundamental irreps of $a_n$ and $c_n$, and
the vector and spinor irreps of $b_n$ and $d_n$. Only for one
algebra, $e_8$, is there no such rep.

A sketch of the proof of this is as follows. First, we need to
know the $\g$-rep $X$ in which the right-hand side of (\ref{Y3})
acts. Since the $J_a$ form an adjoint representation of $\g$, it
is clear from (\ref{Y3'}) that the left-hand side of (\ref{Y3}) is
contained in $\Lambda^2(\g)$, the antisymmetric part of $\g\otimes
\g$. Further (and, as we saw above, equivalent to $H^2(\g)=0$),
\be f_{d(ab}u_{c)d}=0 \hspace{0.2in} \Rightarrow \hspace{0.2in}
u_{ab} = f_{abc}v_c \,, \ee so that $X\oplus \g=\Lambda^2(\g)$,
and it turns out that $X$ is irreducible for all $\g$. The image
of the right-hand side of (\ref{Y3}) in $End(V)$ is zero if
$X\otimes V\not\supset V$  by the Wigner-Eckhart theorem. Knowing
$X$, Drinfeld was then able to find the $V$, listed above, for
which this is true.

\subsubsection{The $Y(\g)$-rep $\g\oplus{\mathbb C}$}

The more general case, in which the $Y(\g)$-irrep is
$\g$-reducible, is much harder. Indeed, the only explicitly known
such rep is again due to Drinfeld, with $V=\g\oplus{\mathbb C}$
(that is, {\em adjoint}$\oplus${\em singlet}). (Note that there
can be no rep of $Y(\g)$ based on the adjoint rep of $\g$ alone,
since $X \subset \g\otimes \g$.) The action of $Y(\g)$ is
\begin{eqnarray}
\nonumber \rho(I_a) x = [I_a,x] &  & \rho(J_a) x = <x,I_a> \\
\label{gplusC}\label{adj} \rho(I_a) \lambda = 0 &  & \rho(J_a)
\lambda = d \lambda I_a
\end{eqnarray}
on $(x,\lambda) \in \g \oplus {\mathbb C}$, where $<,>$ is an
inner product on $\g$, and $d \in {\mathbb C}$  is dependent on
$\g$ and on the choice of inner product. In fact, it is rather
intriguing that, for the exceptional algebras, $d$ depends only on
dim$\,\g$, and that there is a uniform formula for the whole of
the exceptional series $a_2,g_2,d_4,f_4,e_6,e_7,e_8$, including
its classical elements. This, alongside the appearance of $X$ in
\cite{deligne,cvitanovic} and the unified $\cR$-matrix structure
of \cite{westbury}, suggests that it might be interesting to
investigate the connection between Yangians and the `magic square'
construction of the exceptional $\g$.

\subsubsection{The tensor product graph}\label{TPG}

There are no explicit constructions of $Y(\g)$ actions on more
general $\g$-reducible $Y(\g)$-irreps. However, we can construct
tensor products of the reps above, and some conclusions may be
drawn. Let us denote $\g$-reps with upper-case letters
$V,W,\ldots$, and $L_\mu\left(Y(\g)\right)$-reps ({\em i.e.\
}acted on by the automorphism (\ref{auto}), and thus carrying a
parameter $\mu$) with lower-case letters $v(\mu),w(\mu),\ldots$.
The essential point which will emerge is this: suppose we first
decompose $v(\mu)$ into $\g$-irreps; of course we then know the
action of the $I_a$ on each. But we also know that the $J_a$ act
in the adjoint rep, and this limits the $\g$-irreps between which
it may have non-trivial action.

Suppose we wish to construct the tensor product $u(\mu/2)\otimes
u(-\mu/2)$ where $u=U$ is an irrep $\rho$ of the form
(\ref{irreps}). The action on $U$ is \be\label{irreps2}
\tilde\rho^{u(\pm \mu/2)}(I_a) = \rho(I_a) \,, \qquad
\tilde\rho^{u(\pm \mu/2)}(J_a) = \pm {\mu\over 2}\rho(I_a) \,, \ee
 so that the action on the tensor product,
constructed using the coproduct (\ref{Y1del},\,\ref{Y2del}), is
\bea \tilde\rho^{u(\mu/2)\otimes u(-\mu/2)}(I_a) & = &\quad
\rho(I_a)\otimes 1
+ 1 \otimes \rho(I_a)\label{act1}\\[0.1in]
\tilde\rho^{u(\mu/2)\otimes u(-\mu/2)}(J_a) & = & -{\mu\over
2}\left(
 1 \otimes \rho(I_a) - \rho(I_a)\otimes 1\right) + {1\over 4}
\left[\rho\otimes\rho(C),1\otimes \rho(I_a) - \rho(I_a)\otimes
1\right],\qquad\quad\label{act2}\eea where we have used the fact
that \be {1\over 2} f_{abc} I_c\otimes I_b = {1\over 4} \left[ C,
1\otimes I_a-I_a\otimes 1\right]\,,\qquad {\rm with} \quad C\equiv
I_d\otimes I_d\,.\ee It is clear from (\ref{act2}) that the $J_a$
act in the adjoint rep of (and which we shall write as) $\g$, and
that they reverse parity. Further, \be C={1\over 2}
\left(\Delta(I_d)^2 - I_d I_d\otimes 1 - 1 \otimes I_d
I_d\right)\,,\ee so that $C$ takes the numerical value ${1\over
2}\left(C_2(W)-2C_2(U)\right)$ on a component $W$ of $U\otimes U$,
where $C_2=I_dI_d$ is the quadratic Casimir operator of $\g$. Thus
the action of $J_a$ from $W$ to $W'$ may be non-trivial only if
$W'\subset \g\otimes W$, if $W'$ and $W$ have opposite parity, and
if $\mu\neq {1\over 4} \left(C_2(W')-C_2(W)\right)$ (for if this
last equality is satisfied the right-hand side of (\ref{act2})
vanishes). However, the action from $W'$ to $W$ will not then
vanish -- and the $Y(\g)$-rep $u(\mu/2)\otimes u(-\mu/2)$,
irreducible for general $\mu$, will be reducible, but not fully
reducible.

Let us look at an example, with $U=\tableau{1}$ the vector rep of
$SO(N)$. Then \be\tableau{1}\otimes \tableau{1} = \tableau{2}
\oplus \tableau{1 1} \oplus 1\,,\ee where we have denoted by
$\tableau{2}$ the traceless rank-two symmetric tensor, by
$\tableau{1 1}$ the rank-two antisymmetric tensor (the adjoint),
and by $1$ the one-dimensional representation. The $Y(\g)$ action
on $u(\mu/2)\otimes u(-\mu/2)$ is most easily described by forming
these components into a graph, \be \tableau{2}
\;\stackrel{1}{\longrightarrow}\; \tableau{1 1}
\;\stackrel{N/2-1}{\longrightarrow}\; 1\,,\ee with a directed edge
from $W$ to $W'$ labelled with ${1\over
4}\left(C_2(W)-C_2(W')\right)$. We see that this is general
irreducible, but, at $\mu=\pm 1$, for example, is reducible (and
note also that both these possibilities, $\tableau{2}$ and
$\tableau{1 1}\,\oplus 1$, are irreps described earlier, in
(\ref{irreps}) and (\ref{gplusC}) respectively). This `tensor
product graph' (TPG) method \cite{TPG} is generally applicable
provided $U\otimes U$ (or indeed $U\otimes V$, with $V\neq U$) has
no multiplicities.

Via (\ref{R}), it also enables us to determine $\cR(\mu)$, which,
since it commutes with $\tilde\rho(I_a)$, (\ref{act1}), must be of
the form \be \cR(\mu) = \sum_{W \subset U \otimes U} \tau_W(\mu)
P_W \ee (where $P_W$ is the projector onto $W\subset U\otimes U$).
Then its commutation with $\tilde\rho(J_a)$, (\ref{act2}), implies
that, for each pair $W,W'$ of connected nodes of the TPG,
\be\label{taus} \tau_{W'} = {\delta-\mu\over \delta+\mu}\,
\tau_W\,,\qquad{\rm where} \quad\delta = {1\over 4}\left(
C_2(W)-C_2(W')\right)\,.\ee For the admissible $U$ of (\ref{rep}),
this system of equations proves to be consistent and determines
$\cR(\mu)$ up to an overall scalar factor; while for all other $U$
the equations are inconsistent. In our $SO(N)$ example above, the
$\cR(\mu)$ is that found in \cite{zams}. We also see that the
$\cR(\mu)$ constructed this way are rational in $\mu$, justifying
to some extent the claim at the end of the last section.

The above technique, applied to solutions of the YBE (the $\cR$),
is known as the `fusion procedure' \cite{KSR}, and may be used
even if the TPG fails. For example: suppose we wish, in the above
example, to calculate the $\cR$ acting in $W\otimes W$ where
$W=\tableau{1 1}\oplus 1$. Now $W\otimes W$ has various
multiplicities, and the TPG fails, but we can use the fact that
$u(1/2)\otimes u(-1/2)$ is reducible --- that $\cR(-1)$ does not
contain $P_{\tableau{2}}$ --- to construct $\cR$ on $W\otimes
W\subset U^{\otimes 4}$; it is \be\label{R4} \cR(-1)\otimes
\cR(-1)\,.\, 1\otimes \cR(\mu-1)\otimes 1\,.\, \cR(\mu)\otimes
\cR(\mu)\,.\, 1\otimes \cR(\mu+1)\otimes 1\,.\ee The essential
point is that the YBE allows the $\cR(-1)\otimes\cR(-1)$ term,
similar to the projector onto $(\tableau{1 1}\oplus 1)^{\otimes
2}\subset \tableau{1}^{\otimes 4}$, to be moved through to the
right of this expression, which can therefore consistently be
restricted to act on $W\otimes W$. (Incidentally, the resulting
$\cR$ \cite{mack91} can also be calculated directly \cite{chari91}
from the $Y(\g)$ action (\ref{gplusC}) on $\g\oplus{\mathbb C}$.)

\subsection{Yangians and Dorey's rule}

As promised, we finish with an aspect of the representation theory
which is rather nicer for $Y(\g)$ than for $\g$.

We denote by $V_i$ the $i$th fundamental $\g$-irrep, with highest
weight $\omega_i$, and by $v_i(\mu)$ the $i$th fundamental
$L_\mu(Y(\g))$-irrep, which, we recall, has $V_i$ as its `top'
component\footnote{That is, `top' according to a partial ordering
of highest weights in terms of the fundamental weights.}.

\subsubsection{Tensor products of fundamental $\g$-irreps}

For representations of $\g$, the Clebsch-Gordan decomposition
tells us when $V_i\otimes V_j \supset \bar{V}_k$, or, better to
say, when \be V_i \otimes V_j \otimes V_k \supset 1\label{CG}\ee
(where we again denote the one-dimensional rep as $1$). The
solutions to this were characterized in terms of weights in a
longstanding conjecture \cite{PRV}, proved relatively recently
\cite{kumar}. When specialized to the fundamental reps this states
that (\ref{CG}) holds if there exists an element $\sigma$ in the
Weyl group (of transformations of the root lattice generated by
the reflections $w_i$ through planes perpendicular to the simple
roots) $W$ of $\g$ such that the dominant weight\footnote{A
dominant weight is one whose coefficients, when it is expressed in
terms of the simple weights, are all positive. They are also,
therefore, the highest weights.} conjugate to $-\omega_j-\sigma
\omega_k$ is $\omega_i$.

\subsubsection{Dorey's rule}\label{Dorey's rule}

A rule due to Dorey \cite{dorey}, and shown by Braden
\cite{braden} to be a restricted case of the above, was originally
discovered in the context of purely elastic scattering theories
(PESTs, indeed) in 1+1D integrable models, where it described
particle fusings (and thereby $S$-matrix pole structure and
perturbative three-point couplings in affine Toda field theory, at
least when $\g$ is simply-laced \cite{BCDS}). It states that a
fusing $ijk$ occurs iff there exist integers $r$ and $s$ such that
\be \omega_i + c^r \omega_j + c^s \omega_k=0\,,\label{dorey}\ee
where $c=\prod_{i=1}^rw_i$ is a Coxeter element of the Weyl group.
(The ordering of the simple roots does not matter.) The simplest
case in which (\ref{dorey}) is satisfied but (\ref{CG}) is not is
the self-coupling of the second fundamental rep (the rank-two
antisymmetric tensor) of $d_5$.

Coxeter elements have many nice properties which make
(\ref{dorey}) more attractive than (\ref{CG}). Their action
partitions the roots of $\g$ into $r$ orbits, each of size $h$,
where $h$ is the Coxeter number of $\g$ (which therefore satisfies
$r(h+1)=$dim$\,\g$). Further, $c$ acts very simply on $r$ planes
through the root lattice, in each of which it is a rotation by
$s\pi/h$, where $s$ is an exponent of $\g$. In PESTs this yields a
beautiful\footnote{To be convinced of this one only need look at
the illustrations in \cite{coxeter}.} geometric interpretation of
the conservation in three-point couplings of a set of local
charges, with spins equal to the exponents: instead of $r$
conservation equations, one has a single equation in ${\mathbb
R}^r$, projected onto $r$ planes \cite{dorey}.

\subsubsection{Tensor products of fundamental $Y(\g)$-irreps}

Various results from integrable models suggested that Dorey's rule
might apply to the fusion of fundamental $Y(\g)$ reps
\cite{nankai}, and it was proved in 1995 \cite{chari95} that there
exist $\mu,\nu,\lambda$ such that \be v_i(\mu)\otimes v_j(\nu)
\otimes v_k(\lambda) \supset 1\,,\label{CGY}\ee precisely when
(\ref{dorey}) holds --- the values of  $\mu,\nu,\lambda$ are
related to the angles of the rotations described in the last
paragraph. (This is strictly true only for simply-laced $\g$; for
nonsimply-laced $\g$ there is a correspondence, but it is much
more subtle.) The proof, however, was case-by-case, and it is an
open question whether there is a more natural way to access the
hidden geometry of Yangian representations.

\subsection{Some further reading}

An alternative way to present the Yangian story is to begin with
(\ref{RTT})\footnote{Albeit with `quantum' and `auxiliary' spaces
exchanging roles --- see sect.\ref{qq}.} and an explicit
$R$-matrix and proceed from there to construct
$Y({\mathfrak{gl}}_n)$ and, via an appropriate `quantum
determinant', $Y({\mathfrak{sl}}_n)$ (see \cite{molev} and
references therein). Drinfeld also provided another realization of
$Y(\g)$ in \cite{drinf87}, analogous to a Cartan-Weyl basis, which
is often used to study $Y(\g)$ representation theory, and a set of
polynomials in correspondence with (finite-dimensional) $Y(\g)$
reps which can be used to classify them (allowing one, for
example, to deduce the existence of the fundamental $Y(\g)$ irreps
discussed in this chapter). For connections between these two
approaches see \cite{crampe}; for that between $Y(\g)$ and the
Bethe ansatz see \cite{KR,KS}; for that with Hecke algebras see
\cite{cher,drinf86}; for that with separation-of-variables
techniques see \cite{skly}. We noted earlier that $Y(\g)$ may be
thought of as a deformation of a polynomial algebra, with
parameter $z$: the analogue of the full loop algebra, with powers
of $z^{-1}$ as well, is the `quantum double' of $Y(\g)$
\cite{Ydouble1,Ydouble2}. Super-Yangians have their origins in
\cite{nazsuper}; some representations were studied in
\cite{RBZhang}. Finally, $Y(\g)$ can also be obtained in the
$q\rightarrow 1$ limit of a $q$-deformed untwisted affine algebra
$U_q(\hat{g})$ in the `spin gradation' (equivalent for
simply-laced $\g$ to the principal gradation), as remarked in
\cite{drinf1,CPbook} and detailed for $\g=a_1$ in \cite{BL2} -- we
give further details in an appendix. Indeed, the structure of the
representation ring of $U_q(\hat{g})$, for $q$ not a root of
unity, is the same as that of $Y(\g)$.

 \vspace{0.2in}
\section{Yangian symmetry in 1+1D bulk field theory}
\setcounter{equation}{0}

\subsection{A Yangian of classical charges}

\subsubsection{Poisson brackets of charges}

Suppose we have a 1+1D field theory with $\g$ symmetry, the
corresponding conserved current being \be j_\mu(t,x) \in
\g\,,\qquad
\partial^\mu j_\mu=0\,.\label{cons}\ee
If this current further satisfies \be\label{cf}
\partial_0 j_1 - \partial_1 j_0 + [j_0,j_1] = 0
\ee -- as happens, for example, when we can write $j_\mu=
g^{-1}\partial_\mu g$ for some $g$, as in nonlinear sigma models
-- then, upon decomposing the current into components
$j_\mu=j_{\mu\,a}t^a$ of the generators $t^a$ of $\g$, the
 charges \be \label{Q0}Q^{(0)}_a =
\int^{\infty}_{-\infty} j_{0\,a} {\,dx} \ee which generate the
$\g$-symmetry are supplemented by conserved non-local charges \be
Q^{(1)}_a  = \int^{\infty}_{-\infty} j_{1\,a} {dx} + {1\over
2}f_{abc}\int^{\infty}_{-\infty} j_{0\,b}(x)
\left(\int_{-\infty}^{x} j_{0\,c}(y) \,{dy}\right) \,{dx}
\,.\label{Q1}\ee

Classically, we can now use the canonical Poisson brackets of the
current components to investigate the algebra of these. This
current algebra is, for the Gross-Neveu model and its
generalizations \cite{deVEM1},
 \be\label{PB1}
\left\{ j_{\mu\,a} (t,x), j_{\nu\,b} (t,y) \right\} = f_{abc}
j_{\sigma\,c} (t,x) \delta(x-y) \,, \hspace{0.2in} {\rm where}
\hspace{0.1in} \sigma=|\mu-\nu| \,; \ee while for the principal
chiral model, which has two such currents
$j^L_\mu=g^{-1}\partial_\mu g$ and $j^R_\mu=-\partial_\mu
g\,g^{-1}$, it is \cite{FT} \bea\left\{ j_{0\,a} (t,x), j_{0\,b}
(t,y)\right\}
& = & f_{abc} j_{0\,c} (t,x) \delta(x-y) \nonumber\\
\left\{ j_{0\,a} (t,x), j_{1\,b} (t,y) \right\} & = & f_{abc}
j_{1\,c} (t,x) \delta(x-y) -
\delta_{ab} {\partial\over{\partial x}}\delta(x-y)\label{PB2} \\
\nonumber\left\{ j_{1\,a} (t,x), j_{1\,b} (t,y) \right\} & = & 0
\,. \eea Clearly (\ref{PB1}) are  straightforward. In contrast,
(\ref{PB2}), in addition to their lack of covariance, involve the
`non-ultralocal' $\delta'$ term which potentially causes ambiguity
in the charge Poisson brackets (and which is the classical
analogue of the Schwinger term inevitable in $[j_0,j_1]$ in a
quantum current algebra --- see {\em e.g.\ }\cite{IZ}). In fact it
has been argued \cite{FY} that the strange form of (\ref{PB2}) is
a result of taking a non-trivial classical limit of a well-behaved
quantum current algebra.

It is clear that $\{Q^{(1)}_a,Q^{(1)}_b\}$ will include cubic
terms in the $j_0$, but not at all clear that these must be
expressible as a sum of terms cubic in $Q^{(0)}$ \cite{gomes}. In
fact they are not so, but those in
$f_{d(ab}\{Q^{(1)}_{c)},Q^{(1)}_d\}$ are, and the charges form a
classical Yangian (\ref{Y1},\,\ref{Y2},\,\ref{Y3}) \cite{mack92}
(as a Poisson Hopf algebra; let us call it $Y_C(\g)$) under the
correspondence $Q^{(0)}_a\leftrightarrow
I_a,\,Q^{(1)}_a\leftrightarrow J_a$ with $\alpha=1$ and
replacement of commutators by Poisson brackets computed using
(\ref{PB1}). If instead we use (\ref{PB2}), we still obtain a
Yangian, but encounter problems due to the non-ultralocal term.
These are potentially at their worst in (\ref{Y3}), but in fact
are removed by the $f_{d(bc}...$ contraction, while the ambiguity
in (\ref{Y2}) is avoided simply by defining
$\int_{-\infty}^\infty$ as $ \lim_{L\rightarrow\infty}
\int_{-L}^L$. The antipode map (\ref{antipode}) is realized by the
$PT$-transformation \be\label{antipode2} s: \, j_{\mu}(x,t)
\mapsto -j_{\mu}(-x,-t)\,.\ee Finally, one can also compute
$\{M,Q^{(0)}_a\}=0=\{M,Q^{(1)}_a\}$, where $M$ is the (sole, in
1+1D) Lorentz boost generator.

\subsubsection{A classical coproduct}

A classical interpretation of the coproduct is provided
\cite{mack92,LP} by splitting space into two regions (positive and
negative $x$, say), each of which would naturally contain just one
of a pair of asymptotically-separate, particle-like `lumps'. The
two components of the co-product correspond to the integrals over
the two regions, and the non-triviality of (\ref{Y2del}) is
connected to the non-locality of the second term of (\ref{Q1}):
the integral over a range, say $x>0$, which includes one particle
can involve a `tail' $y<0$ which includes the other, and the extra
term in (\ref{Y2del}) results. Explicitly,\bea\nonumber Q^{(0)}_a
& = &
 \int^{0}_{-\infty} j_{0\,a}(x) \,dx + \int^{\infty}_{0}
j_{0\,a} (x) \,dx
\\[2pt]\label{a}
& \equiv & Q^{(0)}_{a-} + Q^{(0)}_{a+} \\[0.1in]\nonumber
{\rm and}\qquad Q^{(1)}_a& = &  \int^{0}_{-\infty} j_{1\,a} (x)
\,dx + \int^{\infty}_{0} j_{1\,a} (x) \,dx + {1\over2}f_{abc}
\left\{ \int^{\infty}_{0} j_{0\,b} (x)\int^{x}_{0}j_{0\,c}
 (y)\,dx\,dy  \right. \\\nonumber
& & + \left. \int^{\infty}_{0} j_{0\,b} (x) \int^{0}_{-\infty}j_{0\,c}
 (y)\,dx\,dy + \int^{0}_{-\infty} j_{0\,b} (x) \int^{x}_{-\infty}j_{0\,c}
  (y) \,dx\,dy \right\} \\[2pt]\label{b}
& \equiv & Q^{(1)}_{a-} + Q^{(1)}_{a+} + {1\over2}f_{abc}
\,Q^{(0)}_{c-} \,Q^{(0)}_{b+} \,, \eea and it is obvious how to
define the coproduct. Note that properties
(\ref{coassoc},\ref{hom}) are guaranteed.

\subsubsection{The Lax formalism}

The classical origin of $T(\lambda)$ lies in the Lax formalism
\cite{ZM}. Define\footnote{Our convention is $\epsilon_{01}=1$, $
\eta_{00}=-\eta_{11}=1$.} \be L_{\mu}(t,x;\lambda) =
{1\over{1-\lambda^2}} \left( j_{\mu}(t,x) +\lambda
\epsilon_{\mu}^{\;\;\nu} j_{\nu}(t,x) \right) \qquad (\lambda \in
{\mathbb C}) \,,\label{Lax} \ee for which the  condition \be
\left[
\partial_0 + L_0, \partial_1 + L_1 \right] = 0 \ee is equivalent
 to both (\ref{cons}) and (\ref{cf}), and use it to define
 $T(x,y;\lambda)$ (at time $t$) via
\be\label{dag} \left( \partial_1 +L_1(x;\lambda) \right)
T(x,y;\lambda)=0 \,. \ee  This gives the $Y_C(\g)$ charges in the
form \cite{deVEM1, deVEM2} \be T(\lambda) \equiv
T(\infty,-\infty;\lambda)= {\bf P} \exp \left(
-\int_{-\infty}^\infty L_1(\xi;\lambda)\right)=\exp \left(
\sum_{r=0}^{\infty} \left(-{1\over\lambda}\right)^{r+1}
Q^{(r)}\right) \ee (where {\bf P} denotes equal-time path ordering
and $Q^{(r)}=Q^{(r)}_a t^a$), giving $Y_C(\g)$ \cite{LP,IK} via
(\ref{Tdel}) and
 \be\label{talg} \left\{ T_{ij}(\lambda) ,
T_{kl}(\mu) \right\} = r_{ip,kq}(\lambda,\mu) T_{pk}(\lambda)
T_{ql}(\mu) -  T_{ip}(\lambda)  T_{jq}(\mu)r_{pk,ql}(\lambda,\mu)
 \ee  where \be r_{ij,kl}(\lambda,\mu)
= {t^a_{ij}\, t^a_{kl}\over{ \mu-\lambda}}\,. \ee (Of course this
relation clearly involves a matrix commutator, and it might more
transparently be written
$\{T(\lambda)\stackrel{\otimes}{,}T(\mu)\}=[r(\lambda,\mu),T(\lambda)\otimes
T(\mu)]$, but this could easily lead to confusion with the
different $\otimes$ of the coproduct.) The way in which the
non-ultralocal terms are handled here is a story in itself
\cite{Maillet}.

 All of this can be achieved without
reference to the spectral parameter $\lambda$ by using the
iterative procedure of \cite{BIZZ}, equivalent to an expansion of
the above in powers of $\lambda$. A similar iteration run in
reverse, equivalent to an expansion of $T$ in powers of $\lambda$,
about $\lambda=0$ rather than $\lambda=\infty$ \cite{wu}, gives
charges which form the Yangian double \cite{Ydouble}.

\subsection{The quantum Yangian}

The quantum version of these charges first appears in a 1978 paper
by L\"uscher \cite{Luescher} (who effectively found much of
$Y({\mathfrak so}_n)$ many years in advance of the general
construction). The closure of the commutator of non-local charges
on cubic terms in the local charges was found by de Vega,
Eichenherr and Maillet \cite{deVEM2}, and L\"uscher's paper was
later re-interpreted and generalized by Bernard \cite{bern90}.

\subsubsection{Quantization of charges}\label{qq}

The first issue is to find the operator product expansion (OPE) of
the currents. A theorem of L\"uscher and Bernard
\cite{Luescher,bern90} states that, under the assumptions of a
local conserved current $j_\mu\in\g$, with a covariant and
$PT$-invariant OPE in terms of the current and its derivatives,
and (in Bernard's version) a smooth UV limit which is a $\g^{(1)}$
Kac-Moody current, the leading terms in the equal-time OPE, in
light-cone coordinates, are \bea\label{OPE1} f_{abc}j_{b\pm}
(x) j_{c\pm}(0) & = & {g \over \,x^{\pm}} j_{a\pm}(0)+... \\
{1\over 2} f_{abc} \left(
j_{b+}(x)j_{c-}(0)-j_{b-}(x)j_{c+}(0)\right) & = & -{g\over 4}
\log(M^2 x^+ x^-)\left(
\partial_+j_{a-}(0) - \partial_-j_{a+}(0)\right)+...\,,
\qquad\quad \label{OPE2}\eea where $g$ is a constant, $g={c_A\over
2i\pi}$ to reproduce (\ref{Y1}), and $M$ a mass scale. Note that
the contraction on the left removes Schwinger terms, which will
not affect the argument below.

(In fact L\"uscher's theorem applies under more general
conditions, without any requirement on the UV limit, and gives an
OPE applicable to, for example, more general nonlinear sigma
models on spaces $G/H$. Whether or not the argument below leads to
a conserved nonlocal charge depends on the structure of $H$
\cite{evans04,abdallas}. For Gross-Neveu models and their
generalizations see \cite{hauer}.)

The condition (\ref{cf}), problematic in the quantum model because
of the divergent product term, now makes sense if it is understood
to be normal-ordered. Similarly a quantum analogue of the first
non-local current (whose time component is being integrated in
({\ref{Q1})) can now be defined, as \be\label{J1Q}
j^{(1)}_{a\,\mu} = \lim_{\delta\rightarrow 0} \left(   Z(\delta)
\epsilon_{\mu\nu}j_a^\nu(t,x) + {1\over 2} f_{abc} j_{b\,\mu}(t,x)
\int^{x-\delta} \hskip -10pt \epsilon_{\rho\sigma}
j_c^\rho(t,y)\,dy^\sigma \right)\,.\ee Upon applying the OPE
(\ref{OPE1},\,\ref{OPE2}) we find that this current is finite if
$Z(\delta)={g\over 2} \log\delta +$const., and conserved if
$Z(\delta) ={g\over 2} \log(M\delta)$.

An important point that we have alluded to but not fully explained
is the distinction between `quantum' and `auxiliary' spaces. We
have described the `quantum' Yangian, which appears as a charge
algebra (and its classical limit) in 1+1D physics. In $T(\lambda)$
we introduced an `auxiliary' algebra, generated by the $t^a$ which
generate $\g\ni j_\mu$. The usage of tensor product ($\otimes$)
notation in (\ref{talg}) would lead to a different, auxiliary
coproduct, for the $t^a$. The origin of `quantum groups' in the
quantum inverse scattering method is through the requirement that
this latter, auxiliary algebra be deformed -- for example, the
discovery of $q$-deformed algebras by the lattice quantization of
the Lax pair of the sine-Gordon \cite{KulResh} and affine Toda
\cite{Jimbo2} models. The method of quantization described above
-- a point-splitting regularization -- is in contrast to this. A
lattice approach to (\ref{Lax}) is not especially fruitful, but
does lead neatly to the auxiliary Yangian \cite{Mack95}, in which
the polynomial algebra generated by the $\lambda^nt^a$ has to be
deformed, as one would expect.

The quantum structure of $Y(\g)$ now follows -- the algebra
(\ref{Y1},\,\ref{Y2}) directly (although (\ref{Y3}) is rather
harder), and the coproducts (\ref{Y1del},\,\ref{Y2del}) by
considering the braiding relations among currents and fields
\cite{bern90}.

\subsubsection{The Lorentz boost}

A key point is that, effectively because of the presence of
$\delta$, so that a $2\pi$ rotation in the $(x,it)$ plane now adds
a loop integral around $x$ to $j^{(1)}$ \cite{bern90}, the quantum
Yangian is no longer merely an internal symmetry. Re-introducing
$\hbar$, so far set equal to $1$, we have \be\label{boost}
[M,Q^{(1)}_a]=-{\hbar^2 c_A\over 4i\pi}\,.\ee Thus a boost of
rapidity $\theta$ to a particle state of rapidity $\phi$ -- that
is, of momentum $(m\cosh\phi, m\sinh\phi)$ -- is now our $Y(\g)$
automorphism $L_\mu$ of (\ref{auto}), with $\mu=-{\hbar c_A\over
4i\pi}\theta$. This immediately implies a physical interpretation
for each appearance of $\mu$ in chapter one, and in particular
that the $\cR$ will serve as $S$-matrices\footnote{Actually one
has both to apply a fixed transformation \cite{karowski} and to
multiply by a $\theta$-dependent scalar factor to achieve the
correct properties, but $\cR$ fixes the $\theta$-dependent matrix
structure.} for multiplets in $Y(\g)$-irreps. Equation (\ref{R})
describes conservation of the charge $x$ in scattering processes,
and the dependence of $\cR$ only on the difference $\nu-\mu$ is an
expression of the covariance of the $S$-matrix. A special value of
$\mu$ at which the tensor product of $Y(\g)$ reps is reducible may
correspond to a pole in the $S$-matrix, and then the `fusion
procedure' on $\cR$ is just the bootstrapped $S$-matrix. For
example, in (\ref{R4}), $\mu=-1$ corresponds to a pole at
$\theta={2i\pi\over N-2}$ at which two $i=1$, vector multiplets
have a bound state corresponding to an $i=2$, $\tableau{1 1}\oplus
1$ multiplet, and (\ref{R4}) is basically $S_{22}$ constructed
from $S_{11}$.

\subsection{Local conserved charges}

We can re-write (\ref{cons},\,\ref{cf}) in light-cone coordinates
as \be\label{lce} \partial_+ j_- = - \partial_- j_+ =  {1\over 2}
[j_-,j_+] \,, \ee and this immediately yields local conservation
equations in the form\be\label{curr}
\partial_- {\rm Tr} (j_{+}^m) =
\partial_+ {\rm Tr} (j_{-}^m) = 0\ee
via cyclicity of trace, with the first ($m=2$) example being
energy-momentum. The use of a trace here requires a matrix
definition of $j_\mu$, perhaps in a defining representation of
$\g$ where one exists, but in fact (\ref{curr}) can be generalized
to avoid this: writing \be\label{str}{\rm Tr} (j_{+}^m)= {\rm
STr}(t^{a_1} t^{a_2} \! \ldots t^{a_m}) j_\pm^{a_1} j_\pm^{a_2}
\ldots j_\pm^{a_m},\ee  we note that such a symmetrized trace
$STr$ is automatically an invariant tensor of $\g$. Thus we are
led immediately to associate a conserved charge with any rank-$m$,
totally symmetric, invariant tensor $d_{a_1 a_2 \ldots a_m}$
associated with a Casimir operator \be\label{cas} {\cal C}_m =
d_{a_1 a_2 \ldots a_m} t^{a_1} t^{a_2} \ldots t^{a_m} \ee where
\be\label{dinv} [ {\cal C}_m , t_b ] = 0 \qquad \iff \qquad
d_{c(a_1a_2 \ldots a_{m-1}}f_{a_m)bc} = 0 \, \ee (and as usual $(
\ldots)$ denotes cycles of the enclosed indices). It is then easy
to check that invariance of $d$ ensures the conservation equations
\be \label{gencons} \partial_\mp( \, d_{a_1 a_2 \ldots a_m}
j_\pm^{a_1} j_\pm^{a_2} \ldots j_\pm^{a_m} \, ) = 0 \ . \ee The
corresponding conserved charges will be denoted \be \label{locch}
q_{\pm s} \,=\, \int_{-\infty}^\infty d_{a_1a_2 \ldots a_m} \,
j_\pm^{a_1}(x) j_\pm^{a_2} (x) \ldots j_\pm^{a_m}(x) \,dx \ee and
labelled by their spin $s= m{-}1$ (the Poisson bracket with the
boost generator $M$ is $\{ M , q_{\pm s} \} = \pm s q_{\pm s}$).

It is straightforward to check that, under either (\ref{PB1}) or
(\ref{PB2}), these (Poisson-)commute with the $Y_C(\g)$ charges --
straightforwardly under (\ref{PB1}), but requiring a nice
cancellation between ultralocal and nonultralocal terms when using
(\ref{PB2}).

It is not the case, however, that all these local charges are in
involution ({\em i.e.\ }Poisson-commute). In fact a mutually
commuting set can be constructed only for certain spins $s$, which
turn out to be precisely the exponents of $\g$, with $m=s+1$ the
ranks of the primitive invariant tensors, and repeating modulo the
Coxeter number. For the full story see \cite{EHMM}.

To attempt to quantize these charges, with their products of many
currents evaluated at one point, using point-splitting or lattice
techniques would be hopeless. Instead, the best one has is an
anomaly-counting technique due to Goldschmidt and Witten
{\cite{GW}, which, when it works, guarantees quantum conservation
of a charge of some particular $s$. The technique cannot be
expected to yield results for large $s$, but for every model of
this type believed to be integrable with $Y(\g)$ symmetry, it
works for at least one low-lying value of $s$ \cite{EHMM,EKMY} --
and it is believed that in 1+1D models  conservation of just one
higher-spin charge is sufficient to guarantee integrability
\cite{Parke}.

Thus it is expected that the quantum theory will include both
$Y(\g)$ charges and this second set of local charges
\cite{EHMM,naka}. These latter, of course, are the remnants after
a massive integrable deformation of $W$-algebraic extended
conformal symmetry \cite{zam}, and are well-known in affine Toda
field theories \cite{wils80}; they are precisely the charges
discussed in section \ref{Dorey's rule}. Thus the particle
multiplets in these models, which will be $Y(\g)$-irreps, must
carry unique values of the local charges, whose conservation will
constrain the allowed fusings. In the simplest models (such as the
Gross-Neveu and principal chiral models, but not the more general
symmetric-space sigma models), in which the particle multiplets
are expected\footnote{From the bootstrap structure of the
$S$-matrices, mainly --- described below.} to be associated with
the fundamental $Y(\g)$-irreps, and for simply-laced $\g$, it
inevitably follows that Dorey's rule must describe the tensor
products of $Y(\g)$ representations. Based on such connections,
and making use of the charges $Q^{(0)}_a Q^{(0)}_a$ and $Q^{(0)}_a
Q^{(1)}_a$, Belavin suggested \cite{bel} that the symmetry
underlying affine Toda theories might even be $Y(\g)/\g$.

\subsection{A mathematics $\leftrightarrow$ physics dictionary}

A glossary of the mathematics of chapter one in terms of the
physics of this chapter.

\hspace{-0.2in}\begin{tabular}{ccc} {\bf Mathematics} & $\qquad
\longleftrightarrow$ & {\bf Physics}\\[0.1in]
Chapter 1 & & Chapter 2 \\[0.1in]
$I_a$ && $Q^{(0)}_a$ \\[2pt]
$J_a$ && $Q^{(1)}_a$ \\[4pt]
coproduct && action on 2-particle states \\
which is a homormorphism && which represents the
algebra\\
and is co-associative && and is consistent on 3-particle
states \\[4pt]
co-unit && vacuum state \\[2pt]
antipode && $PT$-transformation \\[4pt]
automorphism $L_\mu$ && Lorentz boost of rapidity ${4\pi\mu\over
\hbar c_A}$\\[2pt]
intertwiner $\cR(\mu)$ && (proportional to the) $S$-matrix\\[2pt]
intertwining relation (\ref{R}) && conservation of charges in
interactions\\[2pt]
Yang-Baxter equation (\ref{YBE}) && consistent factorization of
multiparticle $S$-matrix \\[2pt]
fundamental $Y(\g)$ irrep $v_i(\mu)$ && particle multiplet of
rapidity ${4\pi\mu\over \hbar c_A}$ \\[2pt]
$Y(\g)$ tensor product rule (\ref{CGY}) && three-point coupling or
$S$-matrix `fusing' \\[2pt]
fusion of $R$-matrices, {\em e.g.\ }(\ref{R4}) && the $S$-matrix
`bootstrap'
\end{tabular}

Finally we remark on the last two lines, the $S$-matrix
`bootstrap' programme, in which $S$-matrix poles are interpreted
as particle states and thereby used to deduce more $S$-matrix
elements. Of course this originated in 3+1D QFT \cite{chew} and
then dropped from view in the early 1970s (just as it was spawning
some mystical popular pseudoscience \cite{capra}), but it has had
an entirely successful new life in integrable 1+1D models, both in
the simpler case of purely elastic scattering ({\em i.e.\ }in
which there are no degenerate multiplets) \cite{BCDS,doreynotes}
and in the more complex Yangian case \cite{ogiev86}, where its
closure on a spectrum consisting of fundamental $Y(\g)$-irreps is
effectively a re-statement\footnote{Although with vastly more work
to be done on the analytic structure, especially of the scalar
factors with which we are always free to re-scale the $\cR$ when
turning them into $S$-matrices.} of (\ref{CGY}).

\subsection{Yangians in conformal field theory}

The quantum Yangian we have described persists, classically at
least, when a Wess-Zumino term is added to the relevant (principal
chiral model) action \cite{EHMM2}, and so is still present within
the vastly greater Kac-Moody symmetry at the conformal point. An
expansion of the Kac-Moody current in its modes gives a formal
bilinear, Sugawara-like expression for the nonlocal charge.
However, these currents depend on a reference point ($-\infty$ in
the model on a line, at which the $j_\mu\rightarrow 0$), from
which to define the last integral in (\ref{Q1}), which can no
longer be sent to $\infty$ in the conformal model (all points
being conformally equivalent). Nevertheless a suitable nonlocal
charge can be defined \cite{BMM}, providing a Yangian symmetry
applicable in the massless-scattering approach to a CFT
\cite{zams2}. However, L\"uscher's theorem is an asymptotic
expansion: it does not control higher terms or the CFT Green's
function, and it is still not understood how the CFT
field$\leftrightarrow$ state correspondence relates to
infinite-dimensional $Y(\g)$ reps and form factors \cite{smirnov}.

An alternative version of the Yangian charges, intrinsically
defined on a circle and with $\g=a_{N-1}$, originates in the
Yangian symmetry \cite{HHTBP} of a spin chain \cite{HS} with
long-range, $1/r^2$ interactions of $L$ spin sites, each with $N$
possible states, in a circle. The `spinon description' of the
$L\rightarrow\infty$ WZW CFT may be found in \cite{schout}.

Finally, there is one further way in which conformal invariance
can be achieved. Note that the models of section 2.1 naturally
have classical conformal invariance -- the Virasoro modes
Tr$(j_\pm^2)^n$ are classically conserved as a result of
(\ref{curr}). This is broken in the quantum theory, with a running
coupling proportional to $c_A$ \cite{Z-J}. But $c_A$ vanishes for
certain supergroups, and specifically for that needed in the
AdS-CFT correspondence, leading to super-Yangian symmetry in this
conformal field theory \cite{string,HY,sCFT}. However, we note the
ubiquitous appearance of $c_A$ in the treatment of Yangians in
massive models: they are likely to need a very different treatment
for these conformal, supergroup models.

%\vfill\pagebreak
\subsection{Some further reading}

We are not attempting here to describe Yangian symmetry in spin
chains and lattice models --- Bernard's is a good introduction
\cite{bern92}. We do, however, point out that, in addition to
Heisenberg spin chains and their generalizations and lattice
analogues, various bosonic nonlinear sigma models with or without
WZ-terms or worldsheet supersymmetry \cite{EHMM2,CZ} and
Gross-Neveu models and their generalizations \cite{deVEM1,deVEM2},
Yangians make appearances in the Hubbard model \cite{uglk,KL}, the
Calogero-Sutherland and related models \cite{ahn,BHK}, the
non-linear Schr\"odinger hierarchy \cite{nls}, integrable 2D
quantum gravity \cite{isg} and even carbon nanotubes \cite{KL} and
monopole moduli spaces \cite{mono}, as well as in various
connections with $W$-algebras \cite{BHK,annecy}. An even more
tantalizing connection is Polyakov's observation \cite{polyakov}
that Wilson loops in 2+1D Yang-Mills obey equations similar to
(\ref{cons},\ref{cf}) and thus lead to non-local conserved
currents --- unfortunately we lack the measure on the space of
loops which would enable these to be converted into useful
non-local charges. Indeed, it seems that wherever a Lie group
symmetry is combined with integrability, Yangians are to be found.

 \vspace{0.2in}
\section{Boundary remnants of $Y(\g)$ symmetry}
\setcounter{equation}{0}

\subsection{Boundary conditions and local charges}

How can a boundary be incorporated into field theories with
$Y(\g)$ symmetry without losing integrability? We take as our
starting point the boundary equation of motion for the model on
$-\infty<x\leq 0$, written in terms of the currents. (For a full
treatment of the principal chiral model (PCM) in terms of the
underlying field $g\in G$ see \cite{mackshort}.) This is, in
light-cone coordinates, \be {\rm Tr}(j_+(0) j_-(0)) =
0\,,\qquad{\rm or} \qquad j_{+\,a}(0)j_{-\,a}(0)=0\,.\ee We solve
this with \be\label{BC} j_+(0)=\alpha(j_-(0))\,,\qquad{\rm
or}\qquad j_{+\,a}(0) = \alpha_{ab}j_{-\,b}(0)\,,\ee for some
linear transformation $\alpha$ on $\g$,
$\alpha:\;t^a\mapsto\alpha_{ba}t^b$.

Now let us require that $\alpha$ be such as to leave precisely one
of each pair $q_s+q_{-s}$ or $q_s-q_{-s}$ of local charges
(\ref{locch}) conserved. (For example, for $s=1$ we might expect
the first charge, energy, to be conserved on the half-line, but
not the second, momentum.) This is so, and the charges still
Poisson-commute, if $\alpha$ is an involutive automorphism,
$\alpha^2=1$ \cite{mackshort}. Thus $\alpha$ decomposes $\g$ into
$\g=\h+\m$, where $\h$ is the subalgebra with $\alpha$-eigenvalue
$+1$, and $\m$ the complementary $-1$ eigenspace, and \be
[\h,\h]\subset \h\,,\qquad [\h,\m]\subset \m\,,\qquad
[\m,\m]\subset \h \ee (so that $(\g,\h)$ is a symmetric pair and
G/H a symmetric space, where $H$ is the subgroup generated by
$\h$). Then (\ref{BC}) is equivalent to \be j_0(0)\in\h\,,\qquad
j_1(0)\in\m\,,\ee which we recognise as Dirichlet and Neumann
components of a mixed boundary condition. (In the PCM, one way of
realizing this in terms of the field $g\in G$ is by imposing a
Dirichlet condition restricting $g(0)\in H$.) Then the Neumann
condition is that $j_1=0$ when restricted to $\h$ -- again, see
\cite{mackshort}.)

\subsection{Boundary remnant of Yangian charges}

On the half-line,\be {d\over dt}Q^{(0)} = j_1(0) \ee vanishes only
on $\h$, so the $G$-symmetry is broken to $H$. A similar
calculation for $Q^{(1)}$ gives \be {d\over dt}Q^{(1)} =
j_0(0)+{1\over 2} \left[j_1(0),Q^{(0)}\right]\,, \label{bQ1}\ee
which vanishes neither on $\h$ nor on $\m$. At first it was
thought that this meant that nonlocal charges were not essential
for integrability \cite{mourad}, but it was later noticed that a
modified set of nonlocal charges is conserved \cite{DMS}, as
follows.

We first choose to write $\h$-indices as $i,j,k,..$ and
$\m$-indices as $p,q,r...$. Then the $\m$ components of
(\ref{bQ1}) are \be {d\over dt}Q^{(1)}_p = {1\over 2} f_{pqi}
j_{1\,q}(0) Q^{(0)}_i \,.\ee We then find that, while the
$Q^{(1)}_p$ are not conserved, the modified charges \be
\widetilde{Q}^{(1)}_p \equiv Q^{(1)}_p + {1\over 4}
f_{piq}\left(Q^{(0)}_iQ^{(0)}_q+ Q^{(0)}_qQ^{(0)}_i\right)\ee are
conserved\footnote{Note that symmetry/antisymmetry on $i$ and $q$
do not cause this to vanish, since $i$ and $q$ run over different
sets. We have chosen this form to provide the necessary properties
of the quantum charges: classically there is of course no
distinction between $Q^{(0)}_i Q^{(0)}_q$ and $Q^{(0)}_q
Q^{(0)}_i$.}. It remains to be proved, probably by extending the
methods of \cite{Luescher, bern90,evans04,abdallas}, that these
charges remain conserved in the quantum theory, but, assuming this
to be so, a useful way to write them (with $\hbar=1$) is  \be
\widetilde{Q}^{(1)}_p \equiv Q^{(1)}_p -{1\over
4}\left[\widetilde{C},Q^{(0)}_p\right]\,,\ee where
$\tilde{C}=Q^{(0)}_iQ^{(0)}_i$, the quadratic Casimir of $\g$
restricted to\footnote{This is not quite the same as the quadratic
Casimir of $\h$, an important distinction when $\h$ is non-simple
and contains a $u(1)$ factor \cite{DMS}.} $\h$.

We denote as $Y(\g,\h)$ the subalgebra of $Y(\g)$ generated by the
$Q^{(0)}_a$ and the $\widetilde{Q}^{(1)}_a$ \cite{DMS,mack02}. For
$\g={\mathfrak{sl}}_n$ and $\h=\mathfrak{o}_n$ or
$\h=\mathfrak{sp}_n$, this is the twisted Yangian of \cite{tY},
while for $\h=\mathfrak{sl}_{n-m}\times \mathfrak{sl}_m$ it is the
reflection algebra of \cite{MR}\footnote{The author prefers to use
the term `twisted Yangian' for all the $Y(\g,\h)$, but the
restricted meaning is more usual.}. The key algebraic property of
$Y(\g,\h)$, which fixes the special form of the
$\widetilde{Q}^{(1)}_a$, is that, computing the coproduct as usual
using (\ref{Y1del},\ref{Y2del}), one finds \cite{DMS} that
$\Delta(Y(\g,\h))\subset Y(\g)\otimes Y(\g,\h)$. This property
makes $Y(\g,\h)$ a `coideal subalgebra'. Its significance is that
boundary states form representations of $Y(\g,\h)$ (just as bulk
states form representations of $Y(\g)$) and, just as the usual
coproduct's being a homomorphism (\ref{hom}) enables two-particle
states to represent the correct symmetry algebra, so this property
enables a state consisting of a bulk particle and a boundary to
represent $Y(\g,\h)$.
%\footnote{It has always struck the author,
%when teaching, that a variety of fundamental laws of physics are
%really statements that, when viewing an object from a distance,
%little can be said about its internal structure. Newton's third
%law means that a system of particles may behave mechanically like
%a single particle; Gauss's law tells us that we are not to enquire
%how dense or small is a `point charge'; imposing conformal
%invariance at the outset tells us that points on a string are not
%distinguished. So, here, a boundary cannot be distinguished from a
%boundary with particles close by.}.

The analogue of $\cR$ and its relation (\ref{R}) is the
`reflection'- or $K$-matrix, which satisfies \be\label{K} K(\mu)
L_\mu(x) = L_{-\mu}(x) K(\mu) \qquad \forall x \in Y(\g,\h)\,. \ee
The analogue of the Yang-Baxter equation  (\ref{YBE}) is  the
`reflection equation' or `boundary Yang-Baxter equation' (bYBE)
 \be\label{bYBE} \cR(\nu-\mu)\,.\, 1\otimes
K(\nu)\,.\, \cR(\mu+\nu) \,.\, 1\otimes K(\mu) = 1\otimes K(\mu)
\,.\, \cR(\mu+\nu) \,.\,1\otimes K(\nu)\,.\,\cR(\nu-\mu)\,,\ee
illustrated in fig.2.
\begin{figure}[htb]
\hspace{32mm} \setlength{\unitlength}{1mm}
\begin{picture}(120,40)(15,15)

\put(110,-5){\line(0,1){50}}\put(40,-5){\line(0,1){50}}
\put(110.1,-5){\line(0,1){50}}\put(40.1,-5){\line(0,1){50}}
\put(110,15){\line(-2,-1){43}} \put(110,15){\line(-2,1){30}}
\put(40,30){\line(-2,-1){30}} \put(40,30){\line(-2,1){40}}
\put(110,30){\line(-1,-1){40}} \put(110,30){\line(-1,1){20}}
\put(40,15){\line(-1,-1){25}} \put(40,15){\line(-1,1){37}}
 \put(60,20){\makebox(0,0){=}}
 \put(17,48){\makebox(0,0){\footnotesize$\cR(\nu\!-\!\mu)$}}
\put(21,26){\makebox(0,0){\footnotesize$\cR(\nu\!+\!\mu)$}}
\put(74,2){\makebox(0,0){\footnotesize$\cR(\nu\!-\!\mu)$}}
\put(91,19){\makebox(0,0){\footnotesize$\cR(\nu\!+\!\mu)$}}
\put(45,30){\makebox(0,0){\footnotesize$K(\nu)$}}
\put(45,15){\makebox(0,0){\footnotesize$K(\mu)$}}
\put(115,30){\makebox(0,0){\footnotesize$K(\mu)$}}
\put(115,15){\makebox(0,0){\footnotesize$K(\nu)$}}
 \end{picture}
\vspace{30mm} \caption{The boundary Yang-Baxter equation}
\end{figure}
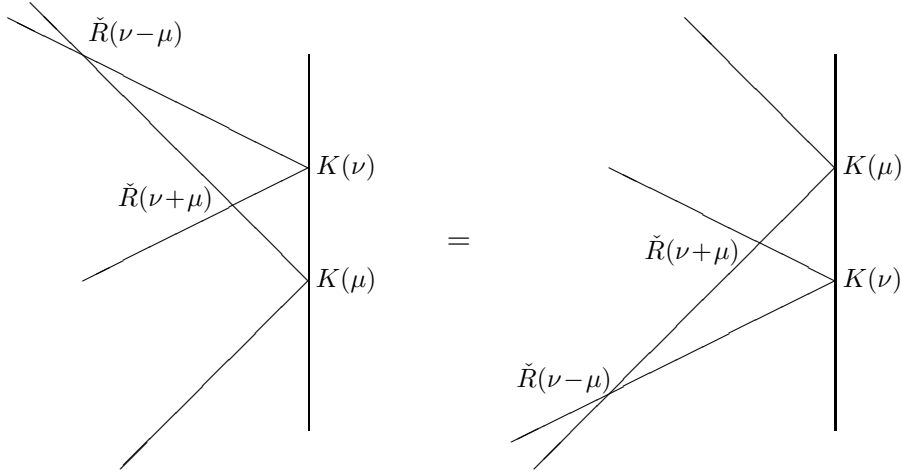
(This can be arrived at through an analogue of the monodromy
matrix $T(\lambda)$, although we do not give the construction
here; the classical $T$ and its relation to the charges $Q^{(i)}$
may be found in \cite{DMS}.) A simple solution, corresponding to
(\ref{firstR}), is \cite{cherbdy} \be\label{firstK}
K(\mu) = \left(\begin{array}{cc} 1 & \mu\\
                      \mu & 1 \\
\end{array}\right) \,.
\ee  More complex solutions may be found in \cite{mackshort}.

\subsection{Representations of $Y(\g,\h)$ and boundary scattering}

Since we used the Yangian charges from the outset of this lecture,
we continue using a physical description here. The use of the bYBE
and the $K$-matrix to describe boundary scattering began with
\cite{cherbdy}; the relationship with boundaries and the monodromy
matrix construction was developed in \cite{sklyanin}; the full
general theory of boundary scattering was worked out in
\cite{ghoshzam}. Analogously to the bulk case, $K(\mu)$ is
(proportional to) the boundary $S$-matrix, (\ref{K}) encodes
conservation of $Y(\g,\h)$ charges in boundary scattering (which
is elastic, so that a particle's rapidity is reversed after
boundary scattering) and  (\ref{bYBE}) is the requirement that
multiparticle boundary scattering factorize consistently.

\subsubsection{The branching graph}

Recall now our general approach to representations of $Y(\g)$ in
section \ref{TPG}. First we decomposed a $Y(\g)$-irrep into its
$\g$-irreducible components, giving us the action of the $I_a$ (or
$Q^{(0)}_a$). Then we used the adjoint $\g$-action of the $J_a$
(or $Q^{(1)}_a$) to deduce their action on these components,
forming them into a graph. A similar approach to $Y(\g,\h)$ begins
by recalling that $Y(\g,\h)\supset\h$, so that its representations
are naturally built on those of the $\h$ generated by the
$Q^{(0)}_i$. Then the $Q^{(1)}_p$ form the rep $\m$ of $\h$
(recall that $[\h,\m]\subset\m$).

Suppose we have a bulk state, a $Y(\g)$-irrep $\tilde\rho$ on
$u(\mu)$, scattering off a boundary. As in section \ref{TPG} let
us consider only the simplest case, in which $u$ is also
$\g$-irreducible, $u=U$,
\be\tilde\rho(Q^{(0)}_a)=\rho(Q^{(0)}_a)\,,\qquad\tilde\rho(Q^{(1)}_a)
= \mu \rho(Q^{(0)}_a)\ee as in (\ref{irreps},\ref{irreps2}). The
intertwiner\footnote{We do not consider here the subtleties of the
cases where boundary scattering also conjugates the state,
$K:\,u\rightarrow u^*$. See for example \cite{doikou}.} $
K_i(\mu): u(\mu) \rightarrow u(-\mu)$ satisfies \be K_v(\mu)
\rho(Q^{(0)}_i) = \rho(Q^{(0)}_i) K_v(\mu) \ee and thus
$K_v(\theta)$ acts trivially on $\h$-irreducible components of
$V$. So we have \be K_v(\mu) = \sum_{W \subset V} \tau_W(\mu)
P_W\,, \ee where the sum is over $\h$-irreps $W$ into which $V$
branches, and $P_W$ is the projector onto $W$. To deduce relations
among the $\tau_W$ we intertwine the $\widetilde{Q}^{(1)}_p$,
using (\ref{irreps2}) analogously to (\ref{act2}), with \be
\tilde\rho(\widetilde{Q}^{(1)}_p) = \mu \rho(Q^{(0)}_p) - {1\over
4}\left[\rho(\widetilde{C}),\rho(Q^{(0)}_p)\right]\,,\ee and
thereby obtain \be \tau_{W'} = {\delta-\mu\over \delta+\mu}\,
\tau_W\,,\qquad{\rm where} \quad\delta = {1\over 4}\left(
\widetilde{C}(W)-\widetilde{C}(W')\right)\,,\ee analogously to
(\ref{taus}), but this time for nodes $W,W'$ (now irreps of $\h$)
connected when $W'\subset \m\otimes W$, in what we might call the
`branching graph'.

As an example let us take, as in section \ref{TPG},
$U=\tableau{1}$ of $SO(N)$, and let $H=SO(M)\times SO(N-M)$. Then
the $N$-dimensional vector $\tableau{1}$ of $\mathfrak{so}_N$
branches to $(\tableau{1},1)\oplus(1,\tableau{1})$ of
$\mathfrak{so}_M\times\mathfrak{so}_{N-M}$, and the graph is
\be\label{bsm}(1,\tableau{1}\,)
\stackrel{(N-2M)/8}{\longrightarrow} (\tableau{1}\,,1)\,.\ee The
directed edge indicates the presence of a pole in the boundary
$S$-matrix at $\mu=-{N-2M\over 8}$ or $\theta={N-2M\over N-2}
{i\pi\over 2}$, and thus a boundary bound state.

Many examples of branching graphs, and a treatment of the more
general, $\g$-reducible case, may be found in \cite{DMS,mack02}.
As with the bulk $S$-matrices, one can conduct a bootstrap
procedure on the boundary $S$-matrices, finding the scattering of
higher bulk particles off the boundary, and using poles like that
in (\ref{bsm}) above to scatter bulk particles off such higher,
non-scalar boundary bound states. Unfortunately the boundary
spectrum seems to be much more complex than the bulk
\cite{shortthesis}, and the procedure has only been completed for
relatively simple cases \cite{short}.

\subsubsection{The symmetric space theorem}

There is an interesting subtlety in the relationship between bulk
and boundary $S$-matrices constructed in this way. In $S$-matrix
theory, the edge of the so-called `physical strip'\footnote{That
is, the physical sheet of the Mandelstam variable $s=(p_1-p_2)^2$,
the (centre-of-mass momentum)$^2$ of the scattering particles.} is
at $\theta=i\pi$ for the bulk $S$-matrix, for which we notice that
a TPG edge directed to the scalar irrep, necessarily from the
adjoint irrep, has label $\mu=c_A/4$ and thus corresponds to an
$S$-matrix pole at $\theta=i\pi$. The edge of the  physical strip
for the boundary $S$-matrix is at $i\pi/2$, and one can deduce
that a branching graph edge directed from $\m$ to the scalar irrep
must correspond to this value; thus $\widetilde{C}(\m)/c_A$ must
equal $1/2$.

We next need to explain the calculation of $\widetilde{C}$.
Suppose $\h$ is non-simple. We first write $ \widetilde{C} =
\sum_i c_i C_2^{\h_i}$, where $\h=\sum_i \h_i$ is a sum of simple
factors $\h_i$ and $C_2^{\h_i}$ is the quadratic Casimir of
$\h_i$. The point here is that $\widetilde{C}$ was written in
terms of generators of $\g$: there will be non-trivial scaling
factors $c_i$, which may be computed by taking the trace of the
adjoint action of $\widetilde{C}$ on $\g$ (where we fix the inner
product to be the identity both on $\g$ and on each $\h_i$),
yielding \be c_i= {c_A \over C_2^{\h_i}(\h_i) + {{\rm dim}\, \m\;
\over {\rm dim}\, \h_i} C_2^{\h_i}(\m)}\,.\ee Then it must be the
case that \be {1\over 2}={\widetilde{C}(\m)\over c_A}={1\over c_A}
\sum_i c_i C_2^{\h_i}(\m)= \sum_i \left( {C_2^{\h_i}(\h_i) \over
C_2^{\h_i}(\m)} + {{\rm dim}\, \m\; \over {\rm dim}\,
\h_i}\right)^{-1}\,.\ee That this holds is a result (discovered in
a very different context) of \cite{godd85}, also known as the
`symmetric space theorem' \cite{daboul96}.

\subsection{Some further reading}

If the thesis of the first two lectures was that $\g$-symmetry
combined with integrability leads to $Y(\g)$-symmetry, at least in
1+1D, then that of this lecture is that an integrable boundary
breaks this to $Y(\g,\h)$ --- but, in contrast to the former case,
evidence for the latter is currently limited to the simple
current-algebra models we have presented. Boundary conditions for
nonlinear sigma models in general seem to have been relatively
little studied --- see \cite{mackyoung} and references therein.
The more general theory of coideal subalgebras may be found in
\cite{letzter}. An introduction to the literature of boundary
integrability for affine Toda theories can be found in
\cite{corr96}.

 Readers may be reminded
by the gluing conditions (\ref{BC}) of the conditions for D-branes
in group manifolds ({\em i.e.\ }the WZNW model on the half-line)
\cite{Db}. Actually there is more freedom in such models, because
the currents there have holomorphic components only --- the
relation is described in \cite{mackshort}.

\vskip 0.4in {\bf Acknowledgments}

This article is based on lectures given as part of the `String
theory in curved backgrounds and boundary conformal field theory'
programme at the Erwin Schr\"odinger Instutute, Vienna in June
2004.

I have benefited from many discussions on these topics over the
years, too many to list here, and I thank everyone for improving
my understanding. I should especially like to thank Vladimir
Drinfeld for a 1991 communication explaining various aspects of
\cite{drinf1}. I should like to thank the ESI for funding my
visit, and the organizers for their hospitality.

\vspace{0.4in}\setcounter{section}{0}

\renewcommand{\thesection}{\Alph{section}}

\section{The Yangian limit of a quantized affine algebra}

Following Jimbo \cite{Jimbo2}, use a Chevalley basis for the
$q$-deformed Lie algebra:
\begin{eqnarray}\nonumber
& \left[ H_a, H_b \right] = 0 & \\ \label{g} & \hspace{1in} \left[
H_a, E^\pm_b \right]\, = \, \left( \alpha_a.\alpha_b \right)
E^\pm_b\, \equiv \, C_{ab} E^\pm_b \hspace{0.3in} {\rm (no \;\;
summation)} &\\ \nonumber & \left[ E^+_a, E^-_a\right] =
{{q^{H_{a}}-q^{-H_{a}}}\over{q-q^{-1}}} &
\end{eqnarray}
where $\alpha_a$, $a=1,...,r$ are the simple roots. This has the
implied normalization \be {\rm Tr}\left( E_a E_{-b} \right) =
\delta_{ab} \hspace{0.3in} {\rm and} \hspace{0.3in} {\rm Tr}
\left( H_a H_b \right) = C_{ab}  \ee (with which the diagonal of
the Cartan matrix $C_{ab}$ is not $2$ but $\alpha_a.\alpha_b\,$.)
Then \beaa
\Delta(H_a) & = & 1\otimes H_a + H_a \otimes 1 \\
\Delta(E^\pm_a) & = &  q^{\pm {H_a/2}} \otimes E^{\pm}_a + E^\pm_a
\otimes q^{\mp {H_a/2}} \,. \eeaa To get the $q$-affine algebra we
append the lowest root $\alpha_0$, and use the `evaluation
automorphism'  in the homogeneous gradation, with all generators
invariant except for \be E^{\pm}_0 \mapsto z^{\pm 1} E^\pm_0,. \ee
But in the Yangian the automorphism (\ref{auto})  affects all
generators equally, so we should apply $x\mapsto UxU^{-1}$ to all
generators, where \be U= z^{\sum_{a=1}^r u_a H_a}
\hspace{0.2in}{\rm and} \hspace{0.2in} h^\vee u_a= \sum_{b=1}^r
C^{-1}_{ab} {2\over\alpha_b.\alpha_b} \,, \ee
 where
$h^\vee$ is the dual Coxeter number. This leads to the `spin
gradation',
\begin{equation}\label{sg}
E^\pm_a \mapsto z^{\pm {2\over\alpha_a.\alpha_a} } E^\pm_a \,,
\end{equation}
for $a=0,1,...,r$. (Note that we could have chosen not to include
the $2/\alpha^2$ factor; we would then, using the Coxeter rather
than dual Coxeter number, have the principal gradation - but we
will need the $2/\alpha^2$ factor later on.)

We now set $q=e^{i\epsilon}$ and $z\equiv q^\mu$, and expand the
generators in powers of $\epsilon$, \be x=x^{(0)} + i \epsilon
\,k(x) x^{(1)} + ... \,, \ee for real constants $k(x)$ to be
determined. (The exchange $E^+ \leftrightarrow E^-$ under
hermitian conjugation forces $\epsilon$ to be real.) Then we fix
the $k$ by requiring the algebra and coproduct of the $x^{(0,1)}$
to be those of the Yangian, and find that, for each $a$, \beaa
H & = & H^{(0)} \\
E^\pm_{\alpha} & = & E^{(0)\pm}_{\alpha} \pm{2\over\alpha.\alpha}
i\epsilon E^{(1)\pm}_{\alpha}\,, \eeaa  where the $H^{(1)}$ must
then be defined by consistency with (\ref{g}). The spin- rather
than principal gradation is needed so that using (\ref{sg}) we
obtain \be E^\pm_{\alpha} \mapsto E^{(0)\pm}_{\alpha}
\pm{2\over\alpha.\alpha} i\epsilon \left( E^{(1)\pm}_{\alpha} +
\mu E^{(0)\pm}_{\alpha} \right) \,, \ee in agreement with
(\ref{auto}).

 \small{

}
\end{document}